\documentclass[preprint]{aastex}

\usepackage{psfig} 

\shortauthors{Rubenstein }

\shorttitle{ Pop~II W~UMas \& Period-Color Rel$^{\rm n}$}

\slugcomment{ This paper has been accepted for publication in the
Astronomical Journal and has been tentatively scheduled for the June
2001 issue.}

\begin{document}

\title{ The effect of stellar evolution upon Population~II Contact
Binaries in the Period-Color Relation.  I. Equal-Mass, Marginal Contact
Systems}

\author{Eric P. Rubenstein\altaffilmark{1}\altaffilmark{2}\altaffilmark{3}} 
\affil{Yale University Astronomy Department, P.O. Box 208101, New
Haven, CT 06520-8101, USA}
\email{ericr@astro.yale.edu}

\altaffiltext{1}{NSF International Research Fellow}
\altaffiltext{2}{J.W. Gibbs Lecturer, Yale University Astronomy Dept.}
\altaffiltext{3}{Visiting post-doctoral fellow at Cerro Tololo Inter-American
Observatory}

\begin{abstract}

Field W~UMa binaries observe a well known Period-Color Relation such
that systems containing more massive stars are bluer and have longer
orbital periods than those systems with lower mass components.
However, it has been known for a decade that metal-poor W~UMa's are
too blue, have too short an orbital period, or both.  Correcting the
observed color for the reduced line blanketing in the atmosphere of a
Pop~II star only accounts for part of the observed discrepancy.  As
others have suggested, and Rucinski (2000) show, the smaller radii of
Pop~II stars and the correspondingly shorter orbital periods are
responsible for the remainder.  In this paper I investigate the effect
of evolution upon the location in the period-color plane.  This paper
addresses the restricted case of equal mass components in critical
contact with their inner Roche lobes, but should be applicable to the
more general cases to the extent that the relative sizes of stellar
components are preserved with metallicity changes.

The calculated metallicity-age dependent Period-Color Relations
substantially agree with Rucinski \& Duerbeck's (1997) empirically
derived corrections to the Period-Color Relation over most of the
investigated range of periods.  However, our predictions deviate to a
greater degree as stellar age increases since their parameterization
does not include the effect of evolution.

\end{abstract}

\keywords{binaries:close---binaries:eclipsing---stars:variables:other---
globular clusters:individual (NGC~6397)}

\section{Introduction}

W Ursae Majoris over-contact binaries (contact binaries here-after)
are comprised of two stars which are sufficiently close that they both
overfill their Roche equipotential surfaces.  They are observed in the
field (e.g. Rucinski 1997, Hendry \& Mochnacki 2000), in open clusters
(Rucinski \& Kaluzny 1994) and in globular clusters (Mateo et
al. 1990; Kaluzny \& Krzeminski 1993; Yan \& Mateo 1994; Rubenstein \&
Bailyn 1996; Kaluzny 1997; Mazur, Krzeminski \& Kaluzny 1999; Mirabal
\& Mateo 1999; Thompson et al. 1999).  In these studies, contact
binaries typically represent roughly a tenth of a percent of the
sample size.  W~UMa's are of great interest to the study of stellar
populations because they can be used as a test of distance (Rucinski
\& Duerbeck 1997, Rubenstein \& Bailyn 1996, Edmonds et al. 1996, Yan
\& Mateo 1994), because even a tiny fraction of binaries can
qualitatively alter the dynamical evolution of dense stellar systems
(Heggie \& Aarseth 1992), and because they are the most frequently
observed type of common envelope objects.

The internal structure of contact binaries is complicated due to such
factors as: the tidal deformation of the components, heat flow and
mass transfer between the stars, possible thermal relaxation
oscillations, chromospheric activity etc. (c.f. Rahunen \& Vilhu 1977;
Webbink 1979; Rahunen 1983; Sarna 1991; Jianmin 1993).  However, many
of their properties can be derived from spectroscopic observations or
via light-curve modeling (see comprehensive text Kallrath \& Milone
1999), especially when the two are used together (Hilditch, King \&
McFarlane 1988).  In particular, the inclination angle ($i$), mass
ratio ($q=M_{2}/M_{1}$) and fill-out ratio (F) can be described by
study of the Fourier components of the light-curve (Rucinski 1993).
Wilson-Devinney (1971 \& 1973) type light-curve fitting (including
recent improvements Kallrath, Milone, Terrell \& Young 1998) allows us
to derive several additional stellar properties such as the surface
temperature of the common envelope, and star-spot sizes, locations,
and temperatures as well as $i$, $q$, \& F (see Kallrath \& Milone
1999 and references within).  As a result of these robust analyses, we
have learned much about the underlying physics of contact systems.
For example, we understand the origin of the Period-Color Relation
(hereafter PCR; Eggen 1967), the observation that W~UMa systems which
are bluer have longer orbital periods.  Briefly, a binary containing a
massive star will be bluer than a binary comprised of lower mass
stars.  Since massive stars are physically larger than low mass stars,
binaries containing a higher mass component must have a greater
orbital separation to accommodate the greater stellar volume within
its Roche lobe.  This larger separation results in a longer period for
these bluer systems.

However, the discovery of W~UMa's in globular clusters (GC's) made it
apparent that the PCR depends upon additional factors.  Yan \& Mateo
(1994) demonstrated that the reduced line blanketing of metal-poor
stellar atmospheres accounted for some of the difference in the GC
W~UMa's positions in the PCR compared to field binaries.  Rucinski
(1995) recognized that metallicity variations will change the stellar
radii and therefore the resulting orbital period of W~UMa systems.
He parameterized
this effect by including a metallicity dependent term and found that
$\Delta {\rm (B-V)} \propto 0.08 \times $ [Fe/H] for metallicity in
the range $-0.5 < {\rm [Fe/H]} < +0.5 $ while for [Fe/H]$<-2$, $\Delta
{\rm (B-V)} \simeq -0.15$.  Rucinski \& Duerbeck (1997) have since
propagated that parameterization to the PCR (their equations 1-3),
though they note that the significant scatter in their Figure~1 arises
from variations in stellar age and the resulting structural changes.
Recently, Rucinski (2000), used the metallicity-radius relation of
Castellani, Degl'Innocenti \& Marconi (1999) to determine the period
changes produced by much lower than solar composition of W~UMa's.
He found that the period decrease was consistent with the
deviation of the population II W~UMa's from the population-I binaries
in the period-color plane.

In \S2 of this work, I will discuss the relationship between the
period and the color of W~UMa systems, taking into account evolution
of the components as well as structural changes arising from
variations in the chemical composition of systems.  I will discuss the
results derived for systems comprised of equal mass components which
are critically filling their inner critical surfaces.  Two example
applications of these metallicity-age dependent Period-Color Relations
are given in \S3.  The more general case of non-equal mass components
and arbitrary fill-out ratios will be discussed in a forthcoming paper
as will the consequences of rapid rotation.

\section{Determining the Metallicity-Age Dependent Period-Color
Relation for non-Solar Composition Binaries}

Rather than try to determine what combination of ``corrections'' would
translate the position of the Pop~II W~UMa's onto the Pop~I
Period-Color Relation (PCR), it is more enlightening to determine what
the location of the PCR should be for a range of ages and
metallicities.  The envelopes of low metallicity stars have lower
opacities than those of higher metallicity stars.  As noted previously
(Yan \& Mateo 1994, Rucinski 1995), this reduced opacity results in a
higher effective temperature, T$_{eff}$, and a bluer color.  However,
the lower opacity also results in a smaller stellar radius for a given
mass and evolutionary state.  Therefore, the components of a low
metallicity contact binary must be closer together than their Pop~I
counterparts in order to maintain a contact configuration.  The
resulting orbital period is shorter.  The combination of these factors
shifts the expected location of W~UMa systems in the Period-Color
plane.  Merely correcting for the reduced line-blanketing of
metal-poor stellar atmospheres does not adequately model this shift
(c.f. Yan \& Mateo 1994 and Rubenstein \& Bailyn 1996).

However, the PCR for a given metallicity can be calculated using the
Roche lobe geometry constraint and stellar models which provide the
radii of the stars of different masses and ages.  Specifically, the
volume of the stars and the volumes of their respective Roche lobes
must be equal for the stars to be in critical contact (both stars
touching the inner Roche potential).  For the case of marginal
contact, a unique period can be calculated for a given pair of masses
and radii (Eggleton 1983):

$$ P_{\rm Orb} = 0.1375\sqrt{\frac{q}{1+q} \mbox{\quad} \frac{1}{r_{L}^{3}} 
\mbox{\quad} \frac{4\pi{\rm R}_{*}^{3}}{3 {\rm M}_{*}^{3}} } $$

where $q={\rm M}_{2}/{\rm M}_{1}$, ${\rm M}_{*}$ and ${\rm R}_{*}$ are
the mass and radius of a component of the binary (Eggleton 1983).  The
dimensionless radius, $r_{L} \sim \frac{0.49q^{2/3}}
{0.6q^{2/3}+ln(1+q^{1/3})}$, and is Eggleton's Eq.~\#2; $r_{L}$ is the
stellar radius divided by the separation of the stars' centers of
mass.  By using models which span a range of masses, the locus of a
PCR can be calculated.  One PCR is derived for each set of age and
composition.

To study the variation of the PCR over a range of age and chemical
composition, I have used stellar models from the Yale Isochrones
(Chaboyer et al.\ 1995; hereafter YI95) to determine the stellar radii
for stars.  In this work I consider stars below the main-sequence
turn-off at solar composition as well as for [Fe/H]=-0.7, -1.0, -1.5,
-2.0 \& -3.0 dex.  I calculated the orbital period expected for cases
where two equal mass stars are both in contact with their inner
critical potential.  This period and the B-V color of the stars (from
YI95) then define the PCR for stars having the same age and
composition.  The curves in the four panels of Figure~\ref{Pcol_byage}
show the PCR for these compositions at 2, 5, 10 \& 15 Gyr.  

How reasonable are the equal mass component and marginal contact
assumptions?  Most W~UMa's are not in critical contact and have mass
ratios significantly less than one.  Certainly the variation of system
properties is important.  However, these two restrictions serve as a
useful starting point.  In any event, many W~UMa's are, in fact,
fairly close to critical contact; in Rucinski \& Duerbeck's (1997,
hereafter RD97) Table~1, 15 out of 33 systems have $f \le 0.15$, {\it
i.e.\ } nearly in critical contact.  To determine how reasonable the
case of critical contact is, let us consider the variation in expected
properties between an idealized system in marginal contact and a
second system with an overcontact of 15\% (i.e. $f = 0.15$).  The 11\%
variation in stellar volume between these two cases corresponds to a
difference in radial separation of less than 4\% (c.f. procedure
detailed in Mochnacki 1984 \& 1985).  The orbital period difference is
then less than 6\%; this factor is negligible compared to the effects
of both metallicity and age.  Furthermore, the important comparison is
between systems of various age/metallicity pairs for a given degree of
contact, so only the {\it differential} change in orbital period at
different age/composition values is relevant and this is an even
smaller effect.  While the bulk of W~UMa's have mass ratios
significantly different from unity, there are exceptions.  RD97's
Table~1 indicates that three out of 33 systems have q$\ge 0.75$.  The
fractional volume difference of two primaries having q=1.0 and 0.75 is
only 6\%, of course the period difference will be larger.  As before,
the procedure delineated above depends upon the differential period
change between systems having different system parameters.  The first
order structural changes are calculated in this work.  The next paper
will correct for these second order effects.

Do solar composition W~UMa's fit these calculated PCR's?  The
asterisks in Figure~\ref{Pcol_byage} mark local W~UMa's from Mochnacki
(1981) and Rucinski (1983) and the pluses mark Rucinski's (1998)
Baade's Window systems.  These calibration systems are plotted in the
two and five Gyr panels only as it is believed that the mean age of
the local W~UMa's is approximately a Gyr (Duerbeck 1984).  Clearly
this application of YI95 models to equal-mass component, contact
configuration binaries adequately fits the calibration data for
systems having approximately solar composition.

What can we learn about metal-poor W~UMa's?  As expected, models of
metal-poor W~UMa's occupy the region further to the blue/with shorter
period than the more metal-rich calibration W~UMa's.  The solid black
dots in each panel correspond to the location in the period-color
plane for specific models (0.5, 0.6 \& 0.7 M$_\odot$, left to right).
These show that at smaller ages, low metallicity W~UMa's have shorter
orbital periods for all masses.  Indeed, even for ages of 15 Gyr,
metal-poor binaries with masses $\lesssim 0.6$ M$_\odot$ have shorter
orbital periods than more metal rich systems.  However, for ages
typical of globular clusters, evolution {\em of those stars at the
turn-off} leads to slightly {\em longer} orbital periods for
critical-contact, metal-poor systems.

We would like to be able to compare the observationally-based results
of RD97 to those expected from these models.  Unfortunately, RD97 and
the preceding papers by Rucinski (1994, 1995 and 1997) do not include
age as a parameter since they are working with calibration objects of
approximately known distance, but unknown age, rather than stellar
models.  Likewise, Rucinski (2000) does not discuss the age of
components as it impacts upon the PCR.  To see what the RD97 relations
might look like with various age stars, the YI95 models are combined
with the RD97 relationships.  The luminosity and stellar volume for a
given mass, age, and metallicity define a particular model.  Using
that model, an inferred orbital period is calculated as above, and the
expected B-V color is determined from RD97's equation \# 1.
Figure~\ref{Pcol_bycomp} shows this comparison of the metallicity-age
dependent (MAD) PCR's and RD97's observationally based
metallicity/color parameterization.

The solar composition models yield nearly identical results for stars
more massive than 0.4 M$_\odot$; the deviations at smaller mass are
presumably due to our limited understanding of stellar opacities for
low temperature stars.  In each panel, the bold lines show the
relevant MAD-PCR while the thin lines show the results of RD97.  The
solid lines are 2 (5) Gyr on the left (right) panels and the dashed
lines are 10 (15) Gyr on the left (right) panels.  The deviations
between the MAD-PCR and RD97 lines are not significant since RD97's
reported $1-\sigma$ uncertainty in magnitude is $\pm0.25$, implying a
color uncertainty of 0.35 mag.  It is therefore apparent that the
structural changes which arise from the reduced opacity of Pop~II
stars accounts for a significant fraction of the PCR differences for
Pop~I and II stars.  Indeed, the shortened orbital period combined
with the reduced line blanketing seem to explain the location of
Pop~II contact binaries, at least given the current limited range of
binary component properties.

\section{Applications}

With many hundreds of field W~UMa's and more than 50 systems in
metal-poor globular clusters (Rucinski 2000 and references therein),
we are strongly motivated to better test the MAD-CPR's against real
data.  As noted above, the bulk of W~UMa's have mass ratios significantly
different from 1; for example, RD97's Table~1 indicates that three out
of 33 systems have q$\ge 0.75$.  At present I will demonstrate the
results of comparing two systems to the B-V MAD-PCR's calculated
above, and a more exhaustive comparison of known W~UMa systems with this
technique will be made after the current model's restrictions are
relaxed.  The planned work will also include V-I MAD-PCR's since many
contact systems are studied in those filters.

In the 15 Gyr panel of figure~\ref{Pcol_byage} the large unfilled
triangle shows the position of V1 in NGC~6397 (Rubenstein \& Bailyn
1996).  V1 falls on the MAD-PCR for [Fe/H]=-2, consistent with the
composition of this globular cluster (Zinn \& West 1984; Djorgovski
1993).  The observational errors are smaller than the size of the
symbol.  The agreement between the metallicity derived from this work
and the cluster's known composition and estimated age ($\approx 18$
Gyr Chaboyer, Demarque \& Sarajedini 1996) is encouraging even though
Rubenstein \& Bailyn's best estimate of V1's mass ratio is $\approx
0.75$, rather than this work's implicit mass ratio of 1.0.

Another indication of the role which future MAD-PCR's might fill
follows.  Covarrubias, Duerbeck \& Maza (2001) found a W~UMa binary
roughly 1.4 kpc above the Galactic plane but were uncertain whether it
is a Pop~I or II object.  This spectroscopically-confirmed binary has
an orbital period of 0.35 day and (B-V)$_o=0.37$.  Inspection of
Figures~\ref{Pcol_byage} \& \ref{Pcol_bycomp} suggest that the object
is likely a Pop~II binary with [Fe/H]$\le -1.0$ dex, however,
[Fe/H]$\approx -0.7$ for an age of only 2 Gyr can not be ruled out on
this basis.

\section{Conclusion}

Metallicity-Age Dependent Period-Color Relations have been calculated
to understand why Population~II contact binaries are observed to be
too blue/have too short an orbital period compared to their Pop~I
analogs.  This work confirms that the structural changes which arise
from the reduced opacity of metal poor stars leads directly to shorter
orbital periods.  The MAD-PCR's calculated agree with the empirically
derived corrections (Rucinski \& Duerbeck 1997) to the Period-Color
Relation to within 1-sigma ($\sim \pm 0.35$ mag in B-V) down to at
least P$_{\rm Orb}\leq 0.2$d.  However, their parameterization does
not include the effect of evolution.  Therefore, the predictions
deviate to a greater degree as stellar age increases.

\begin{figure}[htbp]
\centerline{\psfig{file=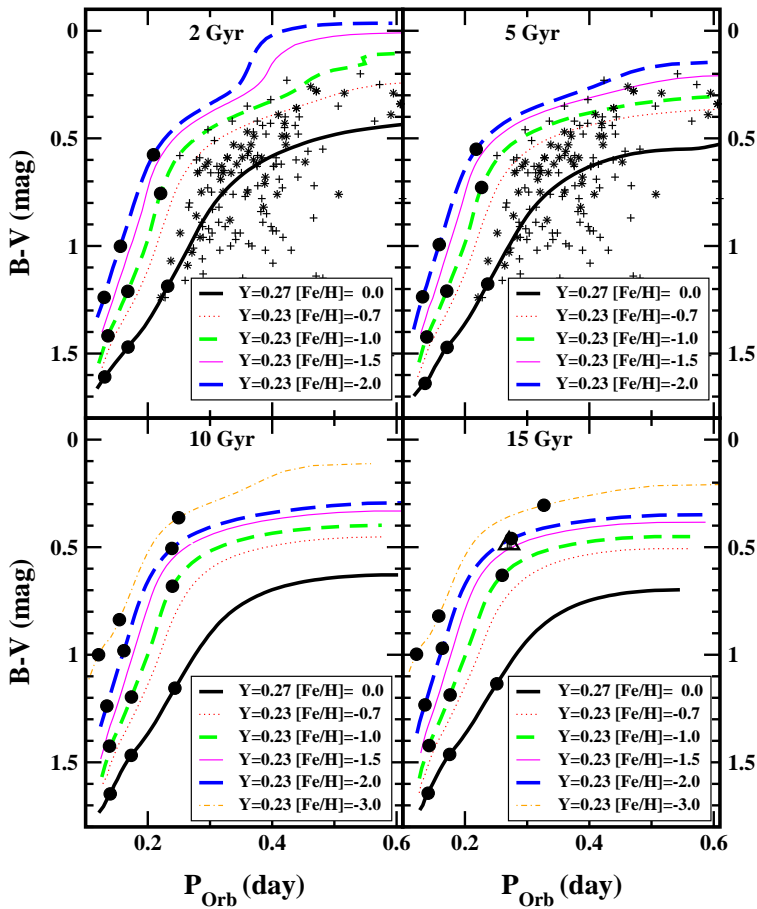,width=4.15in}}
\caption{\small The Metallicity-Age Dependent Period-Color Relation
(MAD-PCR) for contact binaries.  These results refer to equal mass
components which are in critical contact with their inner Roche lobes.
In each panel, corresponding to stars of 2, 5, 10 \& 15 Gyr, the
separate lines refer to MAD-PCR's for solar composition as well as
[Fe/H]=-0.7, -1.0, -1.5, -2.0 (and -3.0 at 10 and 15 Gyr).  The three
solid circles on each curve correspond to the period-color location
for a binary composed of two 0.5, 0.6 \& 0.7 M$_\odot$ stars (from
left to right).  Note that in the upper panels all three of these sets
of models show the expected shortening of the orbital period at lower
metallicity.  The reduced orbital period is due to the smaller opacity
and resulting smaller radii of the stars which leads to a smaller
orbital separation.  Beyond 10 Gyr, the 0.7 M$_\odot$ models show the
effect of the faster evolution of low metallicity stars.  The
resulting increase in stellar radius more than compensates for the
reduced opacity for Pop~II stars and leads to a period increase.  In
the upper panels the asterisks and plus signs correspond to local
W~UMa's (Mochnacki 1981 and Rucinski 1983) and Baade's Window W~UMa's
(Rucinski 1998) respectively.  The open triangle in the 15 Gyr panel
refers to V1 in NGC~6397 from Rubenstein \& Bailyn (1996); it falls on
the [Fe/H]=-2.0, 15 Gyr MAD-PCR.  These quantities are roughly the
same as NGC~6397's composition ([Fe/H]=-1.9 (Zinn \& West 1984;
Djorgovski 1993) and age$\approx 18$ Gyr (Chaboyer, Demarque \&
Sarajedini 1996).  The errors arising from observational uncertainties
are smaller than the size of the triangle.}
\label{Pcol_byage}
\end{figure}

\begin{figure}[htbp]
\centerline{\psfig{file=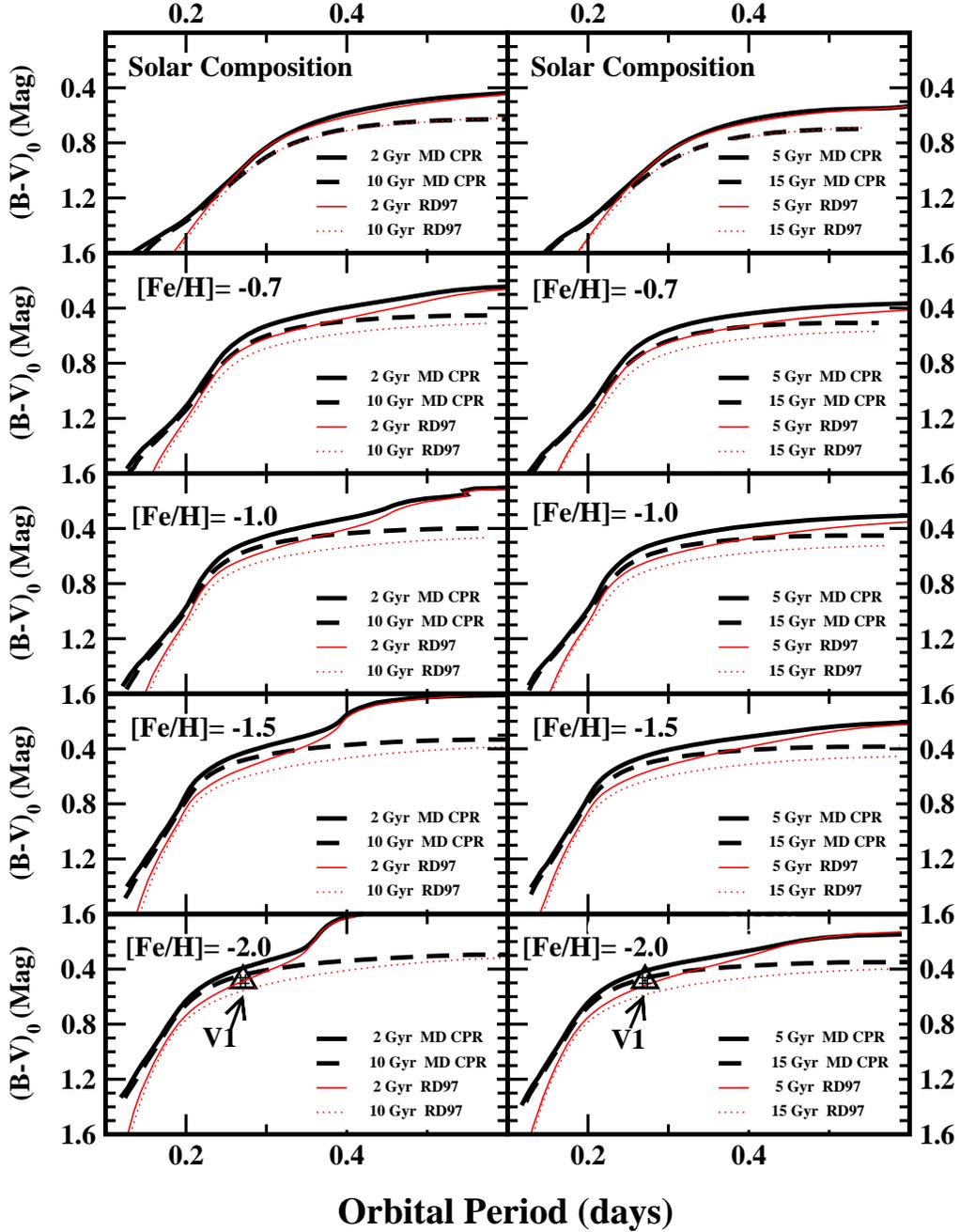,width=5.6in}}
\caption{\small The 10 panels show the MAD-PCR for five different
compositions (solar to [Fe/H]=-2.0, top to bottom).  The left panels
show the calculations for 2 and 10 Gyr, equal mass components in
critical contact with their inner Roche lobes.  The right panels refer
to 5 \& 15 Gyr components.  These calculations agree (to within
$1-\sigma$) with Rucinski \& Duerbeck's (1997; RD97) parameterized
relation for [Fe/H], B-V, and orbital period.  The location of V1 in
NGC~6397 (Rubenstein \& Bailyn 1996) is consistent with the known
properties of its host globular cluster. }
\label{Pcol_bycomp}
\end{figure}

Figures~\ref{Pcol_byage} \& \ref{Pcol_bycomp} show the MAD PCR's for a
range of ages and compositions.  The position of the W~UMa binary
``V1'' from the metal-poor globular cluster NGC~6397 (Rubenstein \&
Bailyn 1996) is shown in the 15 Gyr panel of Figure~\ref{Pcol_byage}
and the [Fe/H]=-2.0 panel of Figure~\ref{Pcol_bycomp}.  This object
lands almost precisely on the MAD PCR appropriate to this cluster.
The success in recovering V1's age and metallicity by looking for a
best fit MAD PCR suggests that this technique may provide an
additional way to derive approximate cluster parameters.  However,
mass transfer may alter the surface abundance of He and metals in the
envelope of the accreting star.  Therefore, any age/metallicity
determinations based on this technique can only be applied to the
cluster as a whole for those W~UMa's which have a mass ratio close to
unity to ensure that little mass transfer contamination has occurred.
Since most W~UMa's do not have mass ratios close to unity (Rucinski \&
Duerbeck 1997), only a small fraction of contact binaries can be used
to obtain age/metallicity values for their host clusters.  In most
instances, this technique would be used to determine the properties of
contact systems for which an age and metallicity were not known (see
\S3).

The planned follow-up work includes extending the range of mass ratios
and degree of contact to more accurately test this technique against
systems with known q \& F.  Additional stellar models having a range
of helium abundance will also be calculated to determine if this
technique is sensitive to the helium enhancement which likely occurs
in the late stages of mass transfer, though structural changes may
make this determination difficult.  In any event, contact systems with
mass ratios close to unity might be able to provide insight into the
relative rates of chemical enrichment of helium and iron among open
clusters with supra-solar abundances.  Finally, models of rapidly
rotating stars will be used to account for changes in stellar radius
and effective temperature.

\acknowledgments

I thank: Alison Sills for information concerning the effects of rapid
rotation; Charles Bailyn and Pierre Demarque for helpful discussions;
Ricardo Covarrubias, Hilmar Duerbeck and Jose Maza for pre-publication
data of their W~UMa; Slavek Rucinski and Stefan Mochnacki for prompt
e-mail responses to related queries and Stefan Mochnacki's code to derive Roche
Lobe parameters; and an anonymous referee for helpful suggestions.  I
also acknowledge the support of NSF grant INT-9902667, the NSF
International Research Fellowship Program, LTSA grant NAGW-2469 and
the hospitality of CTIO where I did most of this work.  This research
has made use of NASA's Astrophysics Data System Bibliographic
Services.

\end{document}